\documentstyle[12pt]{article}
\def\ap#1#2#3{Ann.\ Phys.\ (NY) #1 (19#3) #2}

\def\np#1#2#3{Nucl.\ Phys.\ B#1 (19#3) #2}
\def\pl#1#2#3{Phys.\ Lett.\ #1B (19#3) #2}
\def\pr#1#2#3{Phys.\ Rev.\ D #1 (19#3) #2}
\def\prb#1#2#3{Phys.\ Rev.\ B #1 (19#3) #2}
\def\prep#1#2#3{Phys.\ Rep.\ #1 (19#3) #2}

\def\rmp#1#2#3{Rev.\ Mod.\ Phys.\ #1 (19#3) #2}

\def\cmp#1#2#3{Comm.\ Math.\ Phys.\ #1 (19#3) #2}
\def\cmp#1#2#3{Comm.\ Math.\ Phys.\ #1 (19#3) #2}
\def\nc#1#2#3{Il Nuovo Cimento #1A (19#3) #2}

\newskip\humongous \humongous=0pt plus 1000pt minus 1000pt
\def\caja{\mathsurround=0pt}
\def\eqalign#1{\,\vcenter{\openup1\jot \caja
        \ialign{\strut \hfil$\displaystyle{##}$&$
        \displaystyle{{}##}$\hfil\crcr#1\crcr}}\,}
\newif\ifdtup

\def\frac#1#2{ {{#1} \over {#2} }}
\def\ie{\hbox{\it i.e.}{ }}

\def\half{\mbox{\small $\frac{1}{2}$}}
\def\partder#1{{\partial   \over\partial #1}}
\def\ds#1{\ooalign{$\hfil/\hfil$\crcr$#1$}}
\def\re#1{(\ref{#1})}
\def\beq{\begin{equation}}
\def\eeq{\end{equation}}
\def\beeq{\begin{eqnarray}}
\def\beeqn{\begin{eqnarray*}}
\def\eeeq{\end{eqnarray}}
\def\eeeqn{\end{eqnarray*}}
\def\se{S_{\mbox{\footnotesize{eff}}}}

\def\Vrel{V_{\mbox{\footnotesize{rel}}}}

\def\Grel{\Gamma_{\mbox{\footnotesize{rel}}}}
\def\Irel{I_{\mbox{\footnotesize{rel}}}}
\def\Virr{V_{\mbox{\footnotesize{irr}}}}
\def\Iirr{I_{\mbox{\footnotesize{irr}}}}

\def\si{S_{\mbox{\footnotesize{int}}}}

\def\Girr{\G_{\mbox{\footnotesize{irr}}}}
\def\irr{\mbox{\footnotesize{irr}}}
\def\De{\D_{\mbox{\footnotesize{eff}}}}

\def\DG{\D_{\G}}
\def\DGrel{\D_{\G,\mbox{\footnotesize{rel}}}}

\def\bc{\bar c}

\def\bj{\bar j}
\def\r{\rho}

\def\d{\delta}
\def\de{\delta}
\def\s{\sigma}

\def\S{\Sigma}
\def\G{\Gamma}

\def\L{\Lambda}
\def\l{\lambda}

\def\D{\Delta}
\def\d{\delta}
\def\a{\alpha}
\def\LdL{\L\partial_\L}
\def\UV{$\L_0\to\infty\;$}

\def\bit{\begin{itemize}}
\def\eit{\end{itemize}}
\def\ben{\begin{enumerate}}
\def\een{\end{enumerate}}

\def\nome#1{{\label{#1}}}
\def\p{\partial}

\begin{document}
\begin{titlepage}
\renewcommand{\thefootnote}{\fnsymbol{footnote}}
\begin{flushright}
     UPRF 95-426\\
     LPTHE-95-40\\
     July 1995 \\
\end{flushright}
\par \vskip 10mm
\begin{center}
{\Large \bf
Spontaneous symmetry breaking with \\
Wilson renormalization group}
\end{center}
\par \vskip 2mm
\begin{center}
        {\bf M.\ Bonini} \\
        Dipartimento di Fisica, Universit\`a di Parma and\\
        INFN, Gruppo Collegato di Parma, Italy\\
        and\\
        {\bf M.\ D'Attanasio\footnote{Della Riccia fellow}}\\
        Laboratoire de Physique Th\'eorique et Hautes Energies,
	Universit\'e Pierre et Marie Curie (Paris VI)
	et  Universit\'e Denis Diderot (Paris VII),
	Tour 16, 1er. \'etage, 4, Place Jussieu
	75252 Paris, Cedex 05, France. Laboratoire Associ\'{e} au CNRS URA280
	and\\
        INFN, Gruppo Collegato di Parma, Italy
\end{center}
\par \vskip 2mm
\begin{center} {\large \bf Abstract} \end{center}
\begin{quote}
We study the conditions under which a symmetry is spontaneously broken
in the Wilson renormalization group formulation. Both for a global and
local symmetry, the result is that in perturbation theory one has to
perform a fine tuning of the boundary conditions for the flow of the
relevant couplings. We consider in detail the discrete $Z_2$ case
and the Abelian Higgs model.
\end{quote}
\end{titlepage}

\section{Introduction}
One might think that theories in which one has to deal with an infinite
number of interactions are not predictive.
On the other hand in many cases only a finite number of these
couplings are independent while the others can be expressed in terms of
the independent ones.
This is a general feature whenever one considers the effective theory
of a more fundamental theory in which there are symmetries
which are apparently broken at the effective level, that is at low energies.
A similar situation happens when a (non-anomalous) symmetry of a theory
is broken at the quantum level by the regularization
as for instance in the case of chiral gauge theories \cite{ft}.
In general if there is an infinite number of couplings
one can derive \cite{PW} a set of conditions which constrain them,
preserving renormalizability.
One studies the dependence of the couplings on the ultraviolet cutoff
and seeks solutions of the renormalization  group
equations in which only a finite number of couplings are allowed to be
independent function of the cutoff with the constrain that all
remaining interactions vanish when the independent ones are zero.
It is conjectured that
the solutions of these conditions are the values of the couplings
which restore some symmetry of the theory, hidden by the presence
of an infinity number of interactions.
Then it is also conjectured that in perturbation theory it is sufficient
to restore symmetries at one value of the cutoff after which the
symmetry will be automatically maintained.
This problem of reducing the number of coupling can be analysed in the
Wilson renormalization group (RG) \cite{W}-\cite{B}.
In this formulation one introduces a Wilsonian effective action at a
scale $\L$ in order to take into account the modes above
$\L$. The invariance of the physical Green functions with respect
to variations of $\L$, gives a flow equation for this effective action.

Recently the RG method has been extended to gauge theories
\cite{B}-\cite{others}.
In this case the various couplings do not flow independently.
In particular once the relevant couplings (\ie renormalizable
interactions) are fine tuned at some scale,
the evolution of the Wilsonian action is constrained by the symmetry.
For instance in \cite{B,noi} it is shown that for
the SU(2) Yang-Mills theory there are 9 relevant couplings
which are all expressed in terms of the vector three
point coupling $g$ at some scale (after having fixed the normalization
of the fields) by solving the so called ``fine-tuning'' equations.
Then one prove that the BRS invariance of the physical effective
action is recovered.

A small number of independent couplings is a characteristic feature
also in the case in which the symmetry is spontaneously broken.
In this case the symmetry is broken by the vacuum. In perturbation
theory one does not find terms which break the symmetry so they must be
introduced by hand as interactions and only for a specific value of their
couplings one recovers spontaneous symmetry breaking \cite{PW}.

In this paper we consider the problem of the implementation of a
spontaneously broken symmetry in the RG framework.
One has to distinguish the global symmetry case from the local
one. At first sight the latter is more complicated since the
introduction of a cutoff breaks the symmetry itself. However
due to the Ward identities associated to the local symmetry
this case can be analyzed as the symmetric one.
One introduces the operator which gives the breaking of the Ward
identities due to the presence of the scale $\L$ and studies
its flow with $\L$. The boundary conditions for the relevant part of
this operator determine if one deals with a symmetry which is
spontaneously broken or not. Having chosen one of the two
implementations the flow constraints the couplings in such a way that
the symmetry (spontaneously broken or not) is maintained for any value
of $\L$.

The signature of a spontaneously broken symmetry is the fact that
one scalar field $\phi_0$ acquires a non-vanishing vacuum expectation
value (vev) $\langle\phi_0\rangle$. Thus the theory is symmetric in
the unphysical field $\phi_0$ but it is not in the physical field
$\phi=\phi_0-\langle\phi_0\rangle$.
In this paper we study this property at the effective level.
Namely if the theory is symmetric at any $\L$
in some unphysical field $\phi_0=\phi+v$. We find in general that
in perturbation theory $v$ is $\L$-independent but gets loop
corrections. Besides the field $\phi$ (and then also $\phi_0=\phi+v$)
has a $\L$-dependent value of the vev, which vanishes only for $\L=0$.
In many cases this running of $\langle\phi\rangle$ gives an important
contribution to the running of the masses of the particles.
For instance in
\cite{FP} it was shown how the running of the vev of the dilaton may
be of help in solving the problem of unitarity in some gauge theories
of gravity.

After a brief description of the RG method given in section 2,
we consider in section 3 the case of a scalar theory with a
spontaneously broken $Z_2$ symmetry.
As an example of local symmetries, in section 4 we analyze in detail the
Abelian Higgs model. Section 5 contains some conclusions.

\section{Wilson effective action}
In this section we recall the main features of the Wilson approach for
a simple Euclidean theory with one scalar field, in order to simplify the
notation. The results can be generalized easily to more realistic
theories.

The generating functional is
\beq\nome{Z}
Z[j]=e^{-W[j]}=\int {\cal D}\phi \, \exp{\{-\half (\phi, \,D^{-1}
\phi)_{0\L_0}+(j,\,\phi)_{0\L_0}-\si[\phi;\L_0]\}}
\,,
\eeq
where $D$ is the free propagator of the theory.
We have introduced a cutoff scalar product
$$
(A,\,B)_{\L\L_0}\equiv
\int_p\, K^{-1}_{\L\L_0}(p)\,A(-p)\,B(p)
\,,
\;\;\;\;\;\;\;\;\;\;\;\;\;\;\;\;\;\;
\int_p \equiv \int \frac{d^4p}{(2\pi)^4}\,,
$$
where $K_{\L\L_0}(p)$ is a cutoff function which is one for
$\L^2\le p^2\le\L_0^2$ and rapidly vanishing outside this interval.
$\si[\phi;\L_0]$ is the UV action involving monomials in the fields
and their derivatives which have dimension not larger than four and
are Lorentz scalars.
The Wilsonian effective action $\se$ is defined by integrating over the
energy modes higher than $\L$. One finds
$$
Z[j]=N[j;\L,\L_0] \;\int {\cal D}\phi \,
\exp \{
-\half (\phi,\, D^{-1} \phi)_{0\L}+(j,\,\phi)_{0\L}-\se[\phi;\L,\L_0]
\; \}
\,,
$$
where the coefficient $N$ is given by
\beq\nome{N}
\log N[j;\L,\L_0] = \frac 1 2 (j,\, D j)_{0\L_0}
-\frac 1 2 (j,\, D j)_{0\L} \,.
\eeq
$\se$ contains the effective interactions coming from the
frequencies $p^2>\L^2$. It is easy to show how this functional
is equivalent to a generalization of \re{Z}, in which the free
propagator contains $\L$ as an infrared cutoff. More precisely we have
$$
\se[\phi;\L,\L_0]-\half (\phi,\, D^{-1}\phi)_{\L\L_0} \equiv
W[D^{-1} K^{-1}_{\L\L_0}(p)\phi;\L,\L_0]
\,,
$$
where
$$
e^{-W[j;\L,\L_0]}=\int {\cal D}\phi \,
\exp{\{-\half (\phi, \,D^{-1}\phi)_{\L\L_0}+(j,\,\phi)_{0\L_0}
-\si[\phi;\L_0]\}}\,.
$$
Namely, apart from the tree level two-point function, the Wilsonian
effective action is the generating functional of the connected
Green functions with an IR cutoff $\L$ and amputated of the free
external propagators.
As one expects, it is technically
easier to study the Legendre transform of $W[j;\L,\L_0]$, which is
usually called ``cutoff effective action'' and is a generalization
of the usual quantum effective action, since it contains the infrared
cutoff $\L$ in the free propagators. In the limit $\L\to 0$ and \UV,
one recovers the physical quantum effective action. Both these limits
can be taken in perturbation theory. In particular the dependence on
the ultraviolet cutoff $\L_0$ will be often understood.

{F}rom the fact that the $\L$-dependence of the cutoff effective action
$\G[\phi;\L,\L_0]$ is only coming from the free propagators, one
finds a flow equation in $\L$, the exact RG equation,
which in general has the form
\beq\nome{erg}
\LdL \G[\phi;\L,\L_0]= I[\G;\L,\L_0]\,,
\eeq
where $I$ depends non-linearly on $\G$. For the precise form of $I$
in the various cases see \cite{BDM,WM}.
In order to integrate equation \re{erg} one has to supply
the boundary conditions. For this reason it is useful to split the
cutoff effective action into two parts.
One performs a Taylor expansion of the cutoff vertices at vanishing
momenta (if there are massless particles the expansion should be done
around a non-vanishing subtraction point).
This expansion will have coefficients of decreasing dimension.
These coefficients are the couplings of the theory.
The ``relevant'' part is obtained by  keeping the terms with coefficients
having non-negative dimension (relevant couplings).
The remaining part is called ``irrelevant''.
For instance in the scalar case one gets for the relevant part
$$
\Grel[\phi;\L]=\frac 1 2 \int d^4x \;
\phi(x) [ \s_1(\L)-\s_2(\L) \p^2]
\phi(x) +\frac {\s_3(\L)}{4!} \int d^4x \;
\phi^4(x) \,.
$$
Since we expect the theory to be renormalizable, for $\L\sim\L_0$
the dimension of the irrelevant couplings should be given only by
powers of $\L_0$. Thus the simplest boundary condition for the
irrelevant part of the cutoff effective action is
$$
\Girr[\phi;\L=\L_0]=0\,.
$$
However it would not give any problem to consider irrelevant couplings
which vanish only in the limit \UV.

The boundary conditions for the relevant part fix the physical couplings.
Thus it is natural to set them at the physical point $\L=0$.
For the scalar case this requirement fixes all the three parameters to
be $\s_1(\L=0)=m^2$, $\s_2(\L=0)=1$ and $\s_3(\L=0)=g$.
In the usual field theory language this corresponds to give the
renormalization conditions.
For a more realistic theory the number of relevant couplings
is larger than the physical conditions. This is related to the
fact that some couplings are constrained by the symmetries of the
theory. In general the implementation of a symmetry is equivalent to
a certain number of ``fine tuning'' conditions which fix the
undetermined couplings in terms of the physical ones.
For instance in the SU(2) Yang-Mills case one has $9$ parameters,
of which $3$ are physical conditions (the vector and ghost wave function
normalization and the gauge coupling) and the remaining are given by
$6$ fine tuning conditions \cite{B,noi}.

Once the boundary conditions are fixed, the cutoff effective action
can be obtained by integrating the RG equation
$$
\G[\phi;\L]=\Grel[\phi;\L=0]+
\int_0^\L \frac{d\l}{\l} \Irel +
\int_\L^{\L_0}\frac{d\l}{\l} \Iirr \,,
$$
where the first term provides the boundary conditions for the relevant part,
\ie the physical conditions.
The iterative solution of this integral equation gives the
renormalized loop expansion in terms of the physical couplings
\cite{P,B,BDM}.

\section{Scalar case}
We want to describe the case of a $Z_2$ invariant scalar field
$\phi_0$ with a tree level negative mass parameter $\mu^2$.
This means that $\phi_0$ is not the physical field of the theory and
$\mu^2$ is not the mass of the scalar particle we are describing.
The effective action expressed in terms of the physical field $\phi$
has no longer any $Z_2$ symmetry.
However we impose that the effective action depends on the field
$\phi$ only through the combination $(\phi+v)^2$, where $v$ is
some suitable momentum independent and $\L$-independent quantity,
corresponding to the vacuum expectation value (vev) of the ``unphysical''
field $\phi_0$.

The choice of the boundary conditions for
the relevant part of the cutoff effective action
$$
\Grel[\phi;\L]=\int d^4 x \left\{ \r(\L)\phi(x)
+\frac{1}{2}\phi(x) [\s(\L)-z(\L)\p^2]\phi(x)+
\frac{1}{3!} g_3(\L) \phi^3(x)+
\frac{1}{4!} g(\L) \phi^4(x)
\right\}
$$
must be such that we fix the physical couplings
(the mass $m$ and four-point coupling $g$) and the residual ``reflection''
symmetry corresponding to the spontaneously broken $Z_2$ symmetry.
As expected, this will fix both the value of $g_3=g_3(\L=0)$ and the
vev $v$, in terms of $m$ and $g$.
Indeed, as we shall see, the parameter $v$ is the vev of the field
$\phi_0$ only at $\L=0$, while in general $\langle\phi_0\rangle$ is
$\L$-dependent.
So for the moment we write
\beq\nome{bc}
\Grel[\phi;\L=0]=\int d^4 x \left\{
\frac 1 2 \phi(x) (-\p^2+m^2) \phi(x) +\frac{1}{3!} g_3 \phi^3(x)+
\frac{1}{4!} g \phi^4(x) \right\} \,.
\eeq
Now we come to the determination of $g_3$ and $v$.
It is convenient to consider the cutoff effective potential $V$
\beq
V(\phi;\L)=\sum_{n=1}^\infty \frac{1}{n!} \G_n(\L) \phi^n \,,
\;\;\;\;\;\;\;\;\;\;
\G_n(\L)=\G_n(p_1,\ldots,p_n;\L)\vert_{p_i=0}\,.
\eeq
Notice that also $V$ can be split into its
relevant part $\Vrel=\sum_{n=1}^4 \frac{1}{n!} \G_n(\L) \phi^n$
and irrelevant part
$\Virr=\sum_{n=5}^\infty \frac{1}{n!} \G_n(\L) \phi^n$.
The fact that $V$ is depending on $(\phi+v)^2$, means that
$V(\phi-v;\L)$ is even in $\phi$. Then, by writing
$$
V(\phi-v;\L)=\sum_{n=1}^\infty \frac{1}{n!} G_n(\L) \phi^n \,,
$$
where
$$
G_n(\L)=\sum_{k=n}^\infty \frac{1}{(k-n)!} \G_k(\L) (-v)^{k-n}\,,
$$
we get the relations
$$
G_{2n+1}=0\,.
$$
Notice that the functions $G_n$ are the zero momentum vertices of the
unbroken theory and it holds the relation
$$
G_n(\L)=\frac{\p^n V(\phi;\L)}{\p\phi^n}\vert_{\phi=-v}\,.
$$
We will use the two equations $G_1=0$ and $G_3=0$, namely
\beq\nome{eq}
V'(-v;\L)=0\,,
\;\;\;\;\;\;\;\;\;\;\;\;\;\;\;\;\;
V'''(-v;\L)=0\,,
\eeq
at $\L=0$ to compute the boundary condition $g_3$ and the vev $v$.
In the following we will consider eq.~\re{eq} in perturbation theory.
At the tree level one finds
$$
v^{(0)}=\sqrt{\frac{3 m^2}{g}}\,,
\;\;\;\;\;\;\;\;\;\;\;
g_3^{(0)}=\sqrt{3 m^2 g}\,,
$$
independent of $\L$ and satisfying $g_3^{(0)}=g v^{(0)}$.
At one loop order \re{eq} gives
for the boundary condition of the three point vertex
\beq\nome{res1}
g_3^{(1)}=-\frac{g}{2m^2} {\Virr'}^{(1)}(-v^{(0)};\L=0)
-\frac{1}{4} {\Virr'''}^{(1)}(-v^{(0)};\L=0)
\eeq
and for the vev
\beq\nome{res2}
v^{(1)}=-\frac{1}{2m^2} {\Virr'}^{(1)}(-v^{(0)};\L=0)
+\frac{3}{4g} {\Virr'''}^{(1)}(-v^{(0)};\L=0)\,.
\eeq
Notice that equation \re{res1} gives $g_3$ in terms of irrelevant
vertices of the effective action evaluated at zero momenta.
This is a general feature
and allows one to deduce the perturbative expansion since in the
iterative solution of the RG equations irrelevant vertices al loop
$\ell$ involve relevant couplings at lower loops $\ell'<\ell$.
{F}rom the irrelevant part of the effective potential at one loop
(see appendix A) one finds
\beq\nome{g3}
g_3^{(1)}=-\frac{3 m g \sqrt{3 g}}{64 \pi^2}\,,
\;\;\;\;\;\;\;\;\;\;
v^{(1)}=\frac{9 m\sqrt{3 g}}{64 \pi^2} \,.
\eeq
With this boundary condition for the three point coupling
the one-loop effective potential is completely determined and
it is given by
$$
64 \pi^2 \, V^{(1)}[\phi;\L]=
-\sqrt{3g}m^3 \phi-5g m^2 \phi^2
-3mg\sqrt{3g} \phi^3
-\frac{9 g^2}{8} \phi^4
- \frac{g}{2} \L^2 (\phi+v^{(0)})^2
$$
\beq\nome{V1}
+[(\frac{g}{2} (\phi+v^{(0)})^2 -\frac{m^2}{2} )^2
-\L^4] \log\left(\frac{\L^2+\frac{g}{2}(\phi+v^{(0)})^2-\frac{m^2}{2}}
{m^2}\right)\,.
\eeq
Notice that the one point function vanishes only at $\L=0$ and
is quadratically divergent for large $\L$.
This implies that at any non-vanishing $\L$ the value $\phi=0$ is not
the minimum of the cutoff effective potential. The running minimum
is defined by
$$
\frac{\p V(\phi;\L)}{\p\phi}\vert_{\phi=v(\L)}=0
$$
and at one loop it is given by
$$
v(\L)=-\frac{1}{m^2} \rho(\L)=
\frac{\sqrt{3g} m}{32 \pi^2} \left[-\frac{\L^2}{m^2} +
\log\left(\frac{\L^2+m^2}{m^2}\right) \right]\,.
$$
Thus we have that the running of the one loop vacuum
expectation value of the unphysical field $\phi_0$ is given by
\beq\nome{bareflow}
\langle\phi_0\rangle=v^{(1)}-\frac{1}{m^2} \rho(\L)=
\frac{m\sqrt{3g}}{32 \pi^2}\left[\frac{\L^2}{m^2}+\frac{9}{2}-
\log\left(\frac{\L^2+m^2}{m^2}\right)\right]\,.
\eeq

As well known the loop expansion is insensitive to translations
of the field \cite{CW}.
This implies that at any loop $\ell$ the contribution
to $V^{(\ell)}$ coming from the graphs is the same in the broken and
unbroken theory, namely is a function of $\phi_0^2=(\phi+v^{(0)})^2$.
Thus the graphs contribution in any odd derivative of $V^{(\ell)}$
evaluated at $\phi=-v^{(0)}$ is vanishing. The only thing which
remains in
$\frac{\p^{2n+1}}{\p\phi^{2n+1}}V^{(\ell)}\vert_{\phi=-v^{(0)}}$
is the contribution from the boundary conditions, which is independent
of $\L$. This argument implies that ${V'}^{(1)}(-v^{(0)};\L)$ and
${V'''}^{(1)}(-v^{(0)};\L)$ are independent of $\L$,
giving a $\L$-independent one loop correction to the vev.
This observation holds at any loop order.
At two loops for instance we have from \re{eq}
\beeqn
&&{V'}^{(0)}(-v^{(2)};\L)+{V'}^{(1)}(-v^{(1)};\L)+
{V'}^{(2)}(-v^{(0)};\L)=0\,,
\\
&&{V'''}^{(0)}(-v^{(2)};\L)+
{V'''}^{(1)}(-v^{(1)};\L)+{V'''}^{(2)}(-v^{(0)};\L)=0\,,
\eeeqn
where, for the same reason explained above, $V^{(1)}$ is
even in $\phi+v^{(0)}+v^{(1)}$ and $V^{(2)}$ is
even in $\phi+v^{(0)}$, apart from the boundary conditions.
This justifies in perturbation theory our assumption of
a $\L$-independent value of $v$.

Let us conclude this section with some comments about massless theories.
Because of infrared divergences, some relevant couplings (more
precisely the marginal ones) are defined as the value of the
corresponding vertices at some non-vanishing subtraction points.
At $\L=0$ the various vertices at zero momenta are divergent but the
effective potential is well defined since the scale $\bar \mu$
defining the subtraction points acts as a mass term in the loop integrals.
In massless theories there is the possibility that spontaneous
symmetry breaking is driven by radiative corrections.
This corresponds to having a vanishing tree level value for $g_3$ and
$v$. It is immediate to see that \re{eq} implies that $g_3$
and $v$ vanish at any loop order. Thus it impossible to
have perturbative spontaneous symmetry breaking for a massless
scalar theory. This corresponds to the observation made by
Coleman and Weinberg \cite{CW} that
the non-trivial minimum of the effective potential is out of the validity
of the perturbative expansion. However, there is a possibility to
circumvent this difficulty. Imagine that the boundary condition for
the coupling $g(\L)$ gets loop corrections
\beq\nome{bcstr}
g(\L=0)=g+a \ds{h} g^2\,,
\eeq
where $a$ is some coefficient. It is immediate to check that in this
case the set of equation \re{eq} admits a solution for the
tree level quantities $g_3^{(0)}$ and $v^{(0)}$, which
clearly depends on the scale used for the subtraction point.
A boundary condition like \re{bcstr} is somewhat ``unphysical'', since
now $g$ is no more the value of the physical four point coupling, and
has no clear meaning.
On the other hand if we are dealing with a theory with two or more
fields, in principle it is possible to give a
boundary condition for the self-interaction coupling of one field in a
manner analogous to \re{bcstr}, namely as a series in another coupling,
defined through the second field. This is precisely what was done by
Coleman and Weinberg for the massless scalar QED, in which the four-point
coupling of the scalar is fixed to be zero at tree level and
proportional (with a precise factor) to $e^4$ at one loop, so that a
non-vanishing tree level vev $v$ is obtained\footnote[1]{A numerical
study of a truncation of RG equations for massless scalar QED was
performed in \cite{LTW}.}.
This is an example of how the request of symmetry brings us to make a
``fine tuning'' of one or more boundary conditions.

\section{Abelian gauge symmetry}
In this section we consider the abelian Higgs model in four
dimensional Euclidean space with spontaneously broken
$U(1)$ gauge symmetry. The ``classical action'' for this model is
$$
S=\int d^4 x\biggr\{
\frac 1 4 F_{\mu\nu}F_{\mu\nu} +|D_\mu \phi|^2 -\frac 1 2 m^2
|\phi|^2 + g |\phi|^4 +\frac{1}{2\a} (\p A-\a M \phi_2)^2
+ \bc (\p^2 -\a M^2 - \a e M \phi_1) c
\biggl\}\,,
$$
where $D_\mu=\p_\mu -i e A_\mu$,
$\phi=\frac{1}{\sqrt{2}}(\phi_1+\frac M e +i\phi_2)$, the masses and
couplings are in the relation $2gM^2=e^2 m^2$ and we
have included the 't Hooft gauge fixing term and the ghost action.
This action can be rewritten as
$$
S=\int d^4 x\biggr\{
\frac 1 4 F_{\mu\nu}F_{\mu\nu}+\frac 1 2 M^2 A^2+\frac{1}{2\a}(\p A)^2
+\frac 1 2 |D_\mu \phi_1|^2 +\frac 1 2 m^2 \phi_1^2
$$
\beq\nome{abel2}
+\frac 1 2 |D_\mu \phi_2|^2 +\frac 1 2 \a M^2 \phi_2^2
+\bc(\p^2 -\a M^2)c
\eeq
$$
+\frac g 4 (\phi_1^4+\phi_2^4+2\phi_1^2\phi_2^2+4\frac M e \phi_1\phi_2^2
+4\frac M e \phi_1^3)
+e A_\mu(\phi_2 \p_\mu\phi_1-\phi_1\p_\mu\phi_2)+e M A^2 \phi_1
- \a e M\bc\phi_1 c
\biggl\}\,.
$$
In this form it is easy to see that the free propagators are
$$
D_{\mu\nu}(p)=\frac{1}{p^2+M^2}
(\d_{\mu\nu}-\frac{1-\a}{p^2+\a M^2}p_\mu p_\nu)\,,\;\;\;\;\;
D_{\bc c}(p) =\frac{-1}{p^2+\alpha M^2}\,,
$$
\beq\nome{prop}
D_1=\frac{1}{p^2+m^2}\,,\;\;\;\;\;
D_2=\frac{1}{p^2+\alpha M^2}\,.
\eeq
The action \re{abel2} is invariant under the BRS transformations
\cite{BRS,BRS2}
$$
\d A_\mu=\eta \p_\mu \bc\,,\quad\quad \d\bc =0\,,\quad\quad
\d c=\eta \frac 1 \a (\p A-\a M\phi_2)\,,
$$
$$
\d\phi_1=-\eta e \phi_2 \bc\,,\quad\quad
\d\phi_2=\eta (M\bc+e\phi_1 \bc)\,,
$$
where $\eta$ is a Grassmann parameter.
This invariance is expressed by a set of Slavnov-Taylor (ST)
identities ${\cal S}\,W=0$
for the generating functional $W[J,\chi]$ of the quantum correlation
functions
$$
e^{-W[J,\chi]}=\int{\cal D}\phi\; e^{-S+S_J+S_\chi}\,,
$$
where the source terms are $S_J=\int d^4 x [ j_\mu A_\mu+j_1 \phi_1
+j_2\phi_2+\bj c+\bc j]$ and $S_\chi=\int d^4 x [-\chi_1 e\phi_2 \bc
+\chi_2(M\bc+e\phi_1\bc)]$.
The ST operator is
$$
{\cal S}=(\p_\mu j_\mu)\frac{\d}{\d j}+M\bj\frac{\d}{\d j_2}
-\frac 1 \a \bj\p_\mu \frac{\d}{\d j_\mu}
+j_1 \frac{\d}{\d\chi_1}+j_2 \frac{\d}{\d\chi_2}\,.
$$
For the quantum effective action, defined as
$\G[\Phi,\chi]=W[J,\chi]+S_J$, the ST identities are ${\cal S}_\G\,\G=0$,
where the Slavnov operator is
$$
{\cal S}_\G=(\p_\mu \bc)\frac{\d}{\d A_\mu}
+\frac 1 \a (\p_\mu A_\mu-\a M\phi_2)\frac{\d}{\d c}
-\frac 1 2 \sum_{i=1,2}\left(
\frac{\d\G}{\d\phi_i}\frac{\d}{\d\chi_i}+\frac{\d\G}{\d\chi_i}
\frac{\d}{\d\phi_i}\right)\,.
$$
These expressions can be simplified by noting that the identity
$\bj+\p^2\frac{\d W}{\d j}+\a M \frac{\d W}{\d\chi_2}=0$ holds.
In terms of $\G$ it reads
$$
\frac{\d\G}{\d c}+\p^2 \bc+\a M \frac{\d\G}{\d\chi_2}=0\,.
$$
Using this relation the ST identities become
$$
\bc \p_\mu\frac{\d\G'}{\d A_\mu}
+\sum_{i=1,2}
\frac{\d\G'}{\d\phi_i}\frac{\d\G'}{\d\chi_i}=0\,,
$$
where $\G'=\G-\frac{1}{2\a}(\p_\mu A_\mu-\a M\phi_2)^2$.
The relevant part of $\G$ contains $22$ couplings corresponding to the
renormalizable interactions (see appendix B), whose values are fixed
by the renormalization conditions. We will see that $6$ of them
are given by physical requirements (masses and wave function
normalizations of $A_\mu$ and $\phi_1$, the electric charge $e$ and
the vanishing of the $\phi_2$-$A$ mixing in the two point function)
while the others will be constrained by the symmetry.

We now apply the Wilson method.
By integrating in the path integral the higher momentum modes
$(p^2>\L^2)$ we get
\beq\nome{W}
e^{-W[J,\chi]}=N\;\int{\cal D}\phi \exp\left\{
(-S_2+S_J+\chi_2 M\bc)_{0\L}-\se\right\}\,,
\eeq
where $S_2$ is the part of the action \re{abel2} quadratic
in the fields and the notation $(\cdots)_{0\L}$ stands for the
cutoff scalar product introduced in section~2.
The factor $N$ in \re{W} is given by the product of the four terms
obtained from \re{N} with the substitutions
$j \rightarrow \{j_\mu,\,j_1,\,j_2,\,-\bj\}$ and
$Dj \rightarrow \{D_{\mu\nu}j_\nu,\,D_1 j_1,\,D_2 j_2,\,D_{\bc
c}(j-M\chi_2)\}$.
The functional $\se[\Phi,\chi;\L]$ is the Wilsonian effective
action and is given by
$$
\exp\{-(S_2)_{\L\L_0}-\se[\Phi,\chi;\L]\}
=\int{\cal D}\phi \exp\left\{
(-S_2+\chi_2 M\bc)_{\L\L_0}-\si+S_J\right\}\,,
$$
where the sources in the r.h.s. are related to the field $\Phi$ by
$$
\Phi=K_{\L\L_0}\,(D_{\mu\nu}j_\nu,\,D_1j_1,\,D_2j_2,\,
-D_{\bc c}\bj,\,-D_{\bc c}j+K_{\L\L_0}^{-1}D_{\bc c}M\chi_2)
$$
and $\si$ is the UV action involving monomials in the fields, BRS
sources and their derivatives
which have dimension not higher than four, are Lorentz scalar
and are invariant under charge conjugation.
This, together with the high momentum behaviour of the free
propagators \re{prop} ensures perturbative renormalizability \cite{BDM}.

In the RG formulation one has to define the boundary conditions for
the flow in $\L$ of the Wilsonian action $\se$.
As usual its irrelevant part is fixed at $\L=\L_0$
and vanishes for $\L_0\to\infty$.
The relevant part of $\se$ is fixed at the physical point $\L=0$
in such a way that the the physical effective action fulfills the ST
identities.
In order to show how to fix the parameters of this
functional we first perform the following
change of variables (cutoff BRS) in \re{W}
$$
\d A_\mu=\eta \p_\mu \bc\,,\quad\quad \d\bc=0\,,\quad\quad
\d c=\eta \frac 1 \a (\p A-\a M\phi_2)\,,
$$
$$
\d\phi_1=-\eta K_{0\L} \frac{\d\se}{\d\chi_1}\,,\quad\quad
\d\phi_2=\eta (M\bc-K_{0\L} \frac{\d\se}{\d\chi_2})\,.
$$
One then deduces the following identity
$$
{\cal S} e^{-W}=N\,\int{\cal D}\phi\;\De \;
\exp\left\{(-S_2+S_J+\chi_2 M\bc)_{0\L}-\se\right\}\,,
$$
which gives the violation of the ST identities.
The violation $\De=\D_1+\D_2$ is
\beq\nome{D1}
\D_1=\biggl[ -\phi_1 D_1^{-1} \frac{\d}{\d\chi_1}
-\phi_2 D_2^{-1} \frac{\d}{\d\chi_2}
+(\p_\mu \bc) \frac{\d}{\d A_\mu}
+M\bc \frac{\d}{\d\phi_2}
+\frac 1 \a (\p_\mu A_\mu-\a M\phi_2)\frac{\d}{\d c}
\biggr] \se\,,
\eeq
\beq\nome{D2}
\D_2=\int_p K_{0\L}(p) \sum_{i=1,2} \left(
\frac{\d^2\se}{\d\phi_i \d\chi_i}-
\frac{\d\se}{\d\phi_i}
\frac{\d\se}{\d\chi_i}\right)\,.
\eeq
We now discuss the conditions needed to have $\De=0$, \ie the ST
identities.

The operator $\De$ satisfies a linear evolution equation which
perturbatively has the form
\beq\nome{Dflow}
\LdL\D^{(n)}=L[\D^{(m)}]\,,\;\;\;\;\;\;\;\;\;\;\; m<n\,,
\eeq
where $\D^{(n)}$ is the vertex of $\De$ with $n$ fields and $L$ is a linear
operator. Namely the flow of a vertex of $\De$ is given by the
vertices with lower number of fields. From the linearity of this
equation if one has vanishing boundary conditions then $\De$ is zero
for any $\L$.

In order to discuss the boundary  conditions we distinguish the relevant
part, corresponding to the local approximant with monomials of dimension
not larger than five and with ghost number one (see appendix C),
and the remaining irrelevant part.
{F}rom \re{D1} and \re{D2} the irrelevant part of $\De$ vanishes
at $\L=\L_0\to\infty$ since $\se[\L=\L_0]$ is local and $K_{0\infty}=1$.
We now consider the boundary conditions for the relevant part.
Since we are interested in fixing the relevant couplings of the
effective action at the physical point, the vanishing of the relevant
parameters of $\De$ is imposed at $\L=0$. Moreover at this point the
functional $\De$ is simpler since $\D_2=0$ and $\D_1$ becomes by Legendre
transform the ST functional $\DG\equiv {\cal S}_\G\G$.
As a consequence of the above considerations we have \cite{noi}
$$
\DGrel=0\quad\quad\Rightarrow \quad\quad
\DG=0\,,
$$
where $\DGrel$  is the relevant part of $\DG$.
It is a general feature that the
requirement $\DGrel=0$ gives a number of conditions, the fine tuning
conditions, larger than the number of couplings which have to be fixed.
However due to the nilpotency of the BRS transformation one can find a
set of algebraic consistency conditions which must be identically satisfied
by $\DG$. In our case one can see that
$$
\left[{\cal S}_\G-\frac 1 2 \sum_{i=1,2}\left(
\frac{\d\G}{\d\phi_i}\frac{\d}{\d\chi_i}+\frac{\d\G}{\d\chi_i}
\frac{\d}{\d\phi_i}\right)\right]\DG=0\,.
$$
This is a functional identity which mixes relevant and irrelevant
parts of $\DG$, due to the presence of mass scales (which can be the
masses of the particles, as in our case, or the subtraction point in
a massless theory).
If one shows that these irrelevant parts vanish, the consistency conditions
become a set of relations among the relevant couplings of $\DG$
\cite{BRS2}, thus giving a reduction of the number of independent
relations in $\DGrel=0$.

In general at this point one invokes the so-called
Quantum Action Principle (QAP) which states the locality of $\DG$ at the
first non-trivial loop order.
However one can avoid the use of the QAP and exploit only the
properties of the RG flow. As shown in \cite{noi} the locality of
$\DG$ can be recovered by solving the fine tuning conditions
starting from the vertices of $\DG$ with the lowest number of
fields, \ie $\D^{(2)}$.

We briefly summarize the proof.
Due to the form of the evolution equation \re{Dflow}, perturbatively
the vertex $\D^{(2)}$ does not flow.
Therefore this vertex is given by its relevant part at $\L=0$
and can be set to zero by tuning the couplings in the effective
action.
One then considers the evolution of the vertices $\D^{(3)}$.
{F}rom eq.~\re{Dflow} and the fact that $\D^{(2)}=0$
these vertices do not flow. Therefore they
are equal to their relevant parts at $\L=0$.
By using the consistency conditions one reduces the number of
independent parameters of $\D^{(3)}$ and show that also $\D^{(3)}$
can be set to zero by fixing the couplings of $\G$.
This procedure can be repeated until the whole $\DGrel$ is fixed to zero.
The technique we described is equivalent to the introduction of a
filtration in the analysis of the cohomology of the ST operator
\cite{BBBCD}.

The explicit solution of the various fine tuning equations
is given in appendix D.
After having fixed the following physical conditions
$$
\s_{m_A}=M^2\,,\;\;\;\;\;
\s_A=1-\frac {1} {\alpha}\,,\;\;\;\;\;
\s_{\phi_2A}=0\,,
$$
$$
\s_{m_1}=m^2\,,\;\;\;\;\;\s_1=1\,,\;\;\;\;\;
\s_{A\phi_1\phi_2}=-ie\,,
$$
for the remaining couplings we use the ST identities. One finds
$$
\s_{\bc\chi_2}=M\,,\;\;\;\;\;
\s_{m_2}=\alpha M^2\,,\;\;\;\;\;
\s_2=1\,.
$$
All the other couplings receive perturbative corrections.
As an example we now evaluate the one loop correction to the boundary
condition on the gauge parameter $\s_\a$. This is fixed to be
$$
\s_\a=\frac{1}{\a}+M \frac{\p\S^{(\bc\chi_2)}(p)}{\p p^2}|_{p^2=0}
+iM\frac{\p\S^{(\phi_2 A)}(p)}{\p p^2}|_{p^2=0}\,.
$$
At one loop one finds in the Feynman gauge
$$
\frac{\p\S^{(\bc\chi_2)}(p)}{\p p^2}|_{p^2=0}=
- e^2 M m^2 I_3\,,
$$
$$
\frac{\p\S^{(\phi_2 A)}(p)}{\p p^2}|_{p^2=0}=2 i e^2 M m^2 (I_3-m^2
I_4)+i e^2 \frac{m^4}{M} (I_3-2 m^2 I_4)\,,
$$
where $I_3$ and $I_4$ are given by
$$
16 \pi^2 I_3=\frac{M^2}{(m^2-M^2)^3}\log\frac{M^2}{m^2}+\frac{M^2+m^2}{2 m^2
(m^2-M^2)^2}\,,
$$
$$
16 \pi^2 I_4=\frac{M^2}{(m^2-M^2)^4}\log\frac{M^2}{m^2}+
\frac{5m^2 M^2-M^4+2m^4}{6 m^4(m^2-M^2)^3}\,.
$$
Then the one loop correction to the gauge parameter is
$$
\s_\a^{(1)}=e^2 m^2 [-(m^2+3 M^2) I_3+ 2 m^2 (m^2+M^2) I_4]\,.
$$

\section{Conclusions}

In this paper we have seen what are the boundary conditions that one
has to impose on the RG flow in order to describe a theory with a
spontaneously broken symmetry.
We have discussed the global and the local case.

In the global case, the implementation of the symmetry, which is
immediate in the unbroken phase, requires a non-trivial perturbative
fine tuning of the boundary conditions.
We considered the discrete $Z_2$ symmetry, but the analysis
for a continuous one is identical.

In the local case the implementation of the symmetry
can be performed along the same lines in both phases.
The problem can be reduced to the vanishing of the relevant part
of the Slavnov-Taylor identities.
In principle one could expect that the broken phase is simpler, since
the vector mass avoids problems related to infrared singularities and
there is no need of introducing a non-vanishing subtraction point.
For this reason for instance the extraction of the relevant part of a
functional can be performed at vanishing momentum and reduces to a
truncated Taylor expansion.
In the massless case the non-vanishing subtraction point causes some
technical problems in the solution of the fine tuning equation and the
consistency conditions, since it mixes the relevant and irrelevant
contributions of the various vertices. As a consequence
the filtration of the operator $\DG$ must be introduced.
However we have found that also in the broken phase the solution of
the fine tuning equations is quite complicated. In this case
the mass terms mix the relevant and irrelevant contributions and
a filtration of $\DG$ is also needed.

For simplicity we have considered only the abelian Higgs model.
The non-abelian case can be treated along the same lines.
Obviously one should introduce a non-vanishing subtraction point
in order to define the relevant part of the vertices involving
vector fields of the unbroken symmetry generators.

In the $Z_2$ scalar case we have computed the one loop effective
potential and found that in the direction
in which the symmetry is broken the cutoff effective action acquires a
$\L$-dependent minimum which vanishes only for $\L=0$.
We computed the one-loop value of this running minimum for the scalar
case. The analysis done for
the scalar case (see eqs.~\re{g3}-\re{bareflow}) could be repeated
for the Abelian Higgs model. In this case $v^{(1)}$ is the one
loop correction to the parameter $v^{(0)}=M/e$. The qualitative results
are the same, the only changes are in the factors. That is, at any
non-vanishing $\L$ the value $\phi_1=0$ is not the minimum of the cutoff
effective potential. The running minimum is given by the one-point
function of $\phi_1$.

\noindent
{\bf{Acknowledgements.}} We would like to thank C. Becchi, G. Bottazzi,
for discussions and G. Marchesini and M. Pietroni for
discussions and careful reading of part of the manuscript.
M. D'A. would like to thank the LPTHE (Paris VI-VII) for
kind hospitality.

\vskip 1 true cm
\noindent
{\Large\bf Appendix A}
\vskip 0.3 true cm
\noindent
In this appendix we perform some one loop computations for the
scalar case. The evolution equation for the cutoff effective potential
can be derived from \re{erg}.
However we find more convenient to use the flow equation \cite{HH}
\beq\nome{flow}
\L\frac{\p}{\p\L} V(\phi;\L)=-\frac{\ds{h}}{16\pi^2} \L^4 \log
(\L^2+ V''(\phi;\L))\,.
\eeq
This equation can be obtained from \re{erg} in the
approximation in which the momentum dependent part of the
vertices of the cutoff effective action is discarded (see \cite{AMR}
for the derivation).
By solving this equation iteratively one obtains the loop expansion
of $V$ in this approximation.
Notice that in the one loop case \re{flow} is exact since
in the r.h.s. one must use the tree level value of $V''$
which is momentum independent\footnote[2]{
For the Abelian Higgs model
a formula analogous to \re{flow} gives the exact one loop effective
potential only in the Landau gauge, since in this gauge
the derivative couplings do not contribute to the scalar sector.
}.
In this case it is easy to integrate \re{flow} with the boundary
conditions \re{bc} on the relevant couplings (and vanishing boundary
conditions at $\L=\L_0$ for the irrelevant part).
Thus the \UV limit can be taken and we get
\beq\nome{pot}
64 \pi^2 \, V^{(1)}[\phi;\L]=
-\sqrt{3g}m^3 \phi-5g m^2 \phi^2
+(\frac{64 \pi^2}{6} g_3^{(1)}-\frac{5mg\sqrt{3g}}{2}) \phi^3
-\frac{9 g^2}{8} \phi^4
\eeq
$$
- \frac{g}{2} \L^2 (\phi+v^{(0)})^2
+[(\frac{g}{2} (\phi+v^{(0)})^2 -\frac{m^2}{2} )^2
-\L^4] \log\left(\frac{\L^2+\frac{g}{2}(\phi+v^{(0)})^2-\frac{m^2}{2}}
{m^2}\right)\,,
$$
apart from a field independent term. In this expression the constant
$g_3^{(1)}$ is the boundary condition for the three point coupling and
has to be determined by using \re{res1} in terms of the irrelevant
part $\Virr^{(1)}(\phi;\L=0)$ of this one loop potential at $\L=0$.
In order to isolate this functional one notices that at $\L=0$
the relevant part of \re{pot} is simply given by
$$
\Vrel^{(1)}[\phi;\L=0]=
\frac{1}{6} g_3^{(1)} \phi^3\,,
$$
thus one obtains $\Virr^{(1)}(\phi;\L=0)$ by setting
$g_3^{(1)}=0$ and $\L=0$ in \re{pot}.
By using this result in \re{res1}-\re{res2} one gets \re{g3}.
The effective potential \re{pot} is now completely determined and
given by \re{V1}.
For instance the one loop relevant couplings are
$$
\r^{(1)}(\L)=\frac{\sqrt{3g} m}{32 \pi^2} \left[-\L^2 + m^2
\log\left(\frac{\L^2+m^2}{m^2}\right) \right]\,,
$$
$$
\s^{(1)}(\L)=\frac{g}{32 \pi^2}\left[ -\L^2 -
\frac{3 m^2 \L^2}{\L^2+m^2}
+4 m^2 \log\left(\frac{\L^2+m^2}{m^2}\right)\right]\,,
$$
$$
g_3^{(1)}(\L)=\frac{3mg\sqrt{3g}}{32\pi^2}\left[
-\frac{5\L^4+4 m^2 \L^2 +m^4}{2(\L^2+m^2)^2}+
\log\left(\frac{\L^2+m^2}{m^2}\right)\right]\,,
$$
$$
g^{(1)}(\L)=\frac{3 g^2}{32\pi^2}\left[
-\frac{4 \L^6- m^2 \L^4+ m^4 \L^2}{(\L^2+m^2)^3}+
\log\left(\frac{\L^2+m^2}{m^2}\right)\right]\,.
$$
In the limit $\L\to\infty$ one obtains the one loop values
of the ``bare'' couplings
$$
\r^{(1)}_B=\frac{\sqrt{3g} m}{32 \pi^2} \left(-\L^2 + m^2 T
\right)\,,
\;\;\;\;\;\;\;\;\;\;
\s^{(1)}_B=\frac{g}{32 \pi^2}\left(-\L^2 -3 m^2
+4 m^2 T \right)\,,
$$
$$
g_{3\; B}^{(1)}=\frac{3mg\sqrt{3g}}{32\pi^2}\left(
-\frac{5}{2}+T \right)\,,
\;\;\;\;\;\;\;\;\;\;
g^{(1)}_B=\frac{3g^2}{32\pi^2}\left(
-4+T\right)\,,
$$
where $T=\log(\L^2/m^2)$.
Then the bare potential in terms of the unphysical field $\phi_0$ is
$$
\frac 1 2 \left[-\frac{m^2}{2}+\frac{\ds{h}g}{64\pi^2}
(-2\L^2+3 m^2-m^2 T)\right]\phi_0^2
+\frac{1}{4!}\left[g+\frac{3\ds{h}g^2}{32\pi^2}
(-4+T)\right]\phi_0^4\,.
$$

\vskip 1 true cm
\noindent
{\Large\bf Appendix B}
\vskip 0.3 true cm
\noindent
Here we explicitly give the relevant part of the
effective action $\G[\Phi,\chi]$.
We use the fact that under charge conjugation the functional
$\G[\Phi,\chi]$ is even. Recall that under charge conjugation
$\phi_1$, $\bc$ and $\chi_2$ are even while $A_\mu$, $\phi_2$ and
$\chi_1$ are odd.

Using the equation of motion for the ghost $c$,
the contributions to $\G[\Phi,\chi]$ which contain
relevant couplings are
\beq\nome{rel}\eqalign{
&\G[\Phi,\chi] =
\frac 1 2 \int_p \biggl\{
\G_{\mu\nu}^{(AA)}(p) A_\mu(-p)A_\nu(p)+
\G^{(\phi_1\phi_1)}(p) \phi_1(-p)\phi_1(p)+
\G^{(\phi_2\phi_2)}(p) \phi_2(-p)\phi_2(p)
\cr &+
2\G_\mu^{(\phi_2A)}(p) \phi_2(-p)A_\mu(p)+
2\G^{(\bc\chi_2)}(p) \bc(-p)\chi_2(p)
\biggr\}
\cr &
+ \int_p \int_q
\biggl\{
\frac 1 2 \G_{\mu\nu}^{(2A\phi_1)}(p,q,r) A_\mu(p)A_\nu(q)\phi_1(r)+
\G_\mu^{(A\phi_1\phi_2)}(p,q,r) A_\mu(p)\phi_1(q)\phi_2(r)
\cr &+
\frac {1}{3!}\G^{(3\phi_1)}(p,q,r)\phi_1(p)\phi_1(q)\phi_1(r)+
\frac 1 2 \G^{(\phi_1 2\phi_2)}(p,q,r)\phi_1(p)\phi_2(q)\phi_2(r)
\cr &+
\G^{(\bc\chi_1\phi_2)}(p,q,r) \bc(p)\chi_1(q)\phi_2(r)+
\G^{(\bc\chi_2\phi_1)}(p,q,r) \bc(p)\chi_2(q)\phi_1(r)
\biggr\}
\cr & +
\frac {1}{4!}
\int_p \int_q \int_k
\biggl\{
\G^{(4\phi_1)}(p,q,k,h)\phi_1(p)\phi_1(q)\phi_1(k)\phi_1(h)+
\G^{(4\phi_2)}(p,q,k,h)\phi_2(p)\phi_2(q)\phi_2(k)\phi_2(h)
\cr & +
6\G^{(2\phi_1 2\phi_2)}(p,q,k,h)\phi_1(p)\phi_1(q)\phi_2(k)\phi_2(h)+
6\G_{\mu\nu}^{(2\phi_1 2A)}(p,q,k,h)\phi_1(p)\phi_1(q)A_\mu(k)A_\nu(h)
\cr &+
6\G_{\mu\nu}^{(2\phi_2 2A)}(p,q,k,h)\phi_2(p)\phi_2(q)A_\mu(k)A_\nu(h)+
\G_{\mu\nu\rho\s}^{(4A)}(p,q,k,h)A_\mu(p)A_\nu(q)A_\rho(k)A_\s(h)
\biggr\}
\cr &+\dots\,,
}
\eeq
where $r=-p-q$, $h=-p-q-k$ and the dots stand for
all the remaining terms which contain only irrelevant vertices,
since they are coefficients of monomials in the fields and sources
with dimension higher than four.

The vertices in \re{rel} contain $22$ relevant couplings
which are defined as follows
\beeqn
&&\G_{\mu\nu}^{(AA)}(p)=
\de_{\mu\nu}[\s_{m_A}+p^2 \s_\alpha + \Sigma_L(p)]
+ t_{\mu\nu}(p) [\s_A + \Sigma_T(p)]\,,
\\&&
\G^{(\phi_1\phi_1)}(p)=\s_{m_1}+p^2 \s_1 + \Sigma_1(p)\,,
\\&&
\G^{(\phi_2\phi_2)}(p)=\s_{m_2}+p^2 \s_2 + \Sigma_2(p)\,,
\\&&
\G_\mu^{(\phi_2A)}(p)=p_\mu(\s_{\phi_2A}+ \Sigma^{(\phi_2A)}(p))\,,
\\&&
\G^{(\bc\chi_2)}(p)=\s_{\bc\chi_2}+ \Sigma^{(\bc\chi_2)}(p)\,,
\\&&
\G_{\mu\nu}^{(2A\phi_1)}(p,q,r)=
\de_{\mu\nu}[\s_{2A\phi_1}+\S^{(2A\phi_1)}(p,q,r)]
+\tilde \G_{\mu\nu}^{(2A\phi_1)}(p,q,r)\,,
\\&&
\G_{\mu}^{(A\phi_1\phi_2)}(p,q,r)=
q_\mu[\s_{A\phi_1\phi_2}+\S^{(A\phi_1\phi_2)}(p,q,r)]
+r_\mu[\s_{A\phi_1\phi_2}'+\S^{(A\phi_1\phi_2)}(p,q,r)]\,,
\\&&
\G^{(3\phi_1)}(p,q,r)=
\s_{3\phi_1}+\S^{(3\phi_1)}(p,q,r)\,,
\\&&
\G^{(\phi_1 2\phi_2)}(p,q,r)=
\s_{\phi_1 2\phi_2}+\S^{(\phi_1 2\phi_2)}(p,q,r)\,,
\\&&
\G^{(\bc\chi_1 \phi_2)}(p,q,r)=
\s_{\bc\chi_1 \phi_2}+\S^{(\bc\chi_1 \phi_2)}(p,q,r)\,,
\\&&
\G^{(\bc\chi_2 \phi_1)}(p,q,r)=
\s_{\bc\chi_2 \phi_1}+\S^{(\bc\chi_2 \phi_1)}(p,q,r)\,,
\\&&
\G^{(4\phi_1)}(p,q,k,h)=
\s_{4\phi_1}+\S^{(4\phi_1)}(p,q,k,h)\,,
\\&&
\G^{(4\phi_2)}(p,q,k,h)=
\s_{4\phi_2}+\S^{(4\phi_2)}(p,q,k,h)\,,
\\&&
\G^{(2\phi_12\phi_2)}(p,q,k,h)=
\s_{2\phi_12\phi_2}+\S^{(2\phi_12\phi_2)}(p,q,k,h)\,,
\\&&
\G_{\mu\nu}^{(2\phi_12A)}(p,q,k,h)=
\de_{\mu\nu} [\s_{2\phi_12A}+\S^{(2\phi_12A)}(p,q,k,h)]
+\tilde \G_{\mu\nu}^{(2\phi_12A)}(p,q,k,h)\,,
\\&&
\G_{\mu\nu}^{(2\phi_22A)}(p,q,k,h)=
\de_{\mu\nu} [\s_{2\phi_22A}+\S^{(2\phi_22A)}(p,q,k,h)]
+\tilde \G_{\mu\nu}^{(2\phi_22A)}(p,q,k,h)\,,
\\&&
\G_{\mu\nu\r\s}^{(4A)}(p,q,k,h)=
(\de_{\mu\nu} \de_{\r\s}+\de_{\mu\r} \de_{\nu\s}+\de_{\mu\s}\de_{\nu\r})
[\s_{4A}+\S^{(4A)}(p,q,k,h)]
+\tilde \G_{\mu\nu\r\s}^{(4A)}(p,q,k,h)\,,
\eeeqn
with the conditions
$$
\Sigma_{L}(0)=0\,,
\;\;\;\;\;\;\;\;\;
\frac{\partial \Sigma_{L}(p)}{\partial p^2}|_{p^2=0}=0\,,
\;\;\;\;\;\;\;\;\;
\Sigma_{T}(p)|_{p^2=0}=0\,,
$$
$$
\S_i(0)=0\,,
\;\;\;\;\;\;\;\;\;
\frac{\partial \Sigma_i(p)}{\partial p^2}|_{p^2=0}=0\,,
\;\;\;\;\;\;\;\;\;
i=1,\,2\,,
$$
$$
\S^{(\cdots)}(p_i)|_{p_i=0}=0\,.
$$
In the various $\Sigma$ we can factorize
a dimensional function of $p$. Thus they are irrelevant
and contribute to the irrelevant part of the functional $\G[\Phi,\chi]$.
Similarly the vertices $\tilde \G_i$ are irrelevant since their
Lorentz structure is (partially in the case of $\tilde\G^{(4A)}$)
given by external momenta.

We recall that the ghost propagator and the $\bc$-$c$-$\phi_1$ vertex
are given in terms of the vertices
$\G^{(\bc\chi_2)}$ and $\G^{(\bc\chi_2\phi_1)}$
by
$$
\G^{(\bc c)}(p)=
-p^2 -\alpha M \G^{(\bc\chi_2)}(p)\,,
\;\;\;\;\;\;\;\;\;\;\;\;\;\;\;\;
\G^{(\bc c\phi_1)}=-\alpha M \G^{(\bc\chi_2\phi_1)}\,.
$$
At tree level one has
$$
\G^{(\bc c)}(p)=-(p^2+\alpha M^2)
$$
and
$$
\G^{(\bc c\phi_1)}(p,q,r)=-\alpha M e\,.
$$

\vskip 1 true cm
\noindent
{\Large\bf Appendix C}
\vskip 0.3 true cm
\noindent
We now extract the relevant part of the
most general one dimensional functional of fields and sources
with ghost number $-1$ and odd under charge conjugation.
We call this generic functional $\D$. The vertices of $\D$ which
contain the relevant couplings come from the $\D^{(n)}$ with
$n=2,\dots 5$, where $n$ denotes the number of fields.
{F}rom the two-fields component
$$
\D^{(2)}
=\int_p \biggl\{
\D_\mu^{(\bc A)}(p) \bc(-p)A_\mu(p)+
\D^{(\bc \phi_2)}(p) \bc(-p)\phi_2(p)
 \biggr\}\,,
$$
we have the following relevant parameters
\beeqn
\D_\mu^{(\bc A)}(p)&=&p_\mu [\de_1+ p^2 \de_2+\D_{\irr}^{(\bc A)}(p)]\,,
\\
\D^{(\bc \phi_2)}(p)&=&\de_3+ p^2 \de_4+\D_{\irr}^{(\bc\phi_2)}(p)\,.
\eeeqn
The three-fields component is
$$
\D^{(3)}=\int_p \int_q \biggl\{
\D_\mu^{(\bc A\phi_1)}(p,q,r)\bc(p) A_\mu(q)\phi_1(r) +
\D^{(\bc \phi_1\phi_2)}(p,q,r)\bc(p) \phi_1(q) \phi_2(r)
\biggr\}+\dots\,,
$$
where $r=-p-q$ and the dots stand for
the remaining terms which are all irrelevant.
It contains the following relevant parameters
\beeqn
\D_\mu^{(\bc A\phi_1)}(p,q,r)&=& p_\mu [\de_5+\D_{1,\irr}^{(\bc
A\phi_1)}(p,q,r)]
+ q_\mu [\de_6 +\D_{2,\irr}^{(\bc A\phi_1)}(p,q,r)]\,,
\\
\D^{(\bc \phi_1\phi_2)}(p,q,r)&=&\de_7+q^2\de_8+r^2\de_9+q\cdot r
\de_{10}+\D_{\irr}^{(\bc \phi_1\phi_2)}(p,q,r)\,.
\eeeqn
{F}rom the four-fields component
\beeqn
\D^{(4)}&=&
\int_p \int_q \int_k \biggl\{
\D_{\mu\nu\rho}^{(\bc 3A)}(p,q,k,h)
\bc(p) A_\mu(q) A_\nu(k) A_\rho(h)
\\&+&
\D_{\mu\nu}^{(\bc 2A\phi_2)}(p,q,k,h)
\bc(p) A_\mu(q) A_\nu(k) \phi_2(h)+
\D_{\mu}^{(\bc A2\phi_1)}(p,q,k,h)
\bc(p) A_\mu(q) \phi_1(k) \phi_1(h)
\\&+&
\D_{\mu}^{(\bc A2\phi_2)}(p,q,k,h)
\bc(p) A_\mu(q) \phi_2(k) \phi_2(h)+
\D^{(\bc 3\phi_2)}(p,q,k,h)
\bc(p) \phi_2(q) \phi_2(k) \phi_2(h)
\\&+&
\D^{(\bc 2\phi_1 \phi_2)}(p,q,k,h)
\bc(p) \phi_1(q) \phi_1(k) \phi_2(h)
 \biggr\}+\cdots\,,
\eeeqn
where $h=-p-q-k$, we have the relevant parameters
\beeqn
\D_{\mu\nu\rho}^{(\bc 3A)}(p,q,k,h)&=&(q_\mu \de_{\nu\rho}+
k_\nu \de_{\mu\rho}+h_\rho \de_{\mu\nu})
[\de_{11}+\D_{1,\irr}^{(\bc 3A)}(p,q,k,h)]
\\
&+&[(k+h)_\mu \de_{\nu\rho}+(q+k)_\rho \de_{\mu\nu}+
(q+h)_\nu \de_{\mu\rho}][\de_{12}+\D_{2,\irr}^{(\bc 3A)}(p,q,k,h)]
\\
&+&
\tilde\D_{\mu\nu\rho}^{(\bc 3A)}(p,q,k,h)\,,
\\
\D_{\mu\nu}^{(\bc 2A\phi_2)}(p,q,k,h)&=&\de_{\mu\nu}[\de_{13}+
\D_{\irr}^{(\bc 2A\phi_2)}(p,q,k,h)]+
\tilde\D_{\mu\nu}^{(\bc 2A\phi_2)}(p,q,k,h)\,,
\\
\D_{\mu}^{(\bc A2\phi_1)}(p,q,k,h)&=&
p_\mu [\de_{14}+\D_{1,\irr}^{(\bc A2\phi_1)}(p,q,k,h)]+
q_\mu [\de_{15}+\D_{2,\irr}^{(\bc A2\phi_1)}(p,q,k,h)]
\,,
\\
\D_{\mu}^{(\bc A2\phi_2)}(p,q,k,h)&=&
p_\mu [\de_{16}+ \D_{1,\irr}^{(\bc A2\phi_2)}(p,q,k,h)] +
q_\mu [\de_{17}+ \D_{2,\irr}^{(\bc A2\phi_2)}(p,q,k,h)]\,,
\\
\D^{(\bc 3\phi_2)}(p,q,k,h)&=&\de_{18}+\D_{\irr}^{(\bc 3\phi_2)}(p,q,k,h)
\,,
\\
\D^{(\bc 2\phi_1 \phi_2)}(p,q,k,h)&=&\de_{19}+
\D_{\irr}^{(\bc 2\phi_1 \phi_2)}(p,q,k,h)\,.
\eeeqn
Finally the five-fields component
\beeqn
\D^{(5)}&=&
\int_p \int_q \int_k \int_h
 \biggl\{
\D_{\mu\nu}^{(\bc 2A\phi_1\phi_2)}(p,q,k,h,s)
\bc(p)A_\mu(q) A_\nu(k) \phi_1(h)\phi_2(s)
\\&+&
\D^{(\bc 3\phi_1\phi_2)}(p,q,k,h,s)
\bc(p)\phi_1(q)\phi_1(k)\phi_1(h)\phi_2(s)
\\ &+&
\D^{(\bc \phi_1 3\phi_2)}(p,q,k,h,s)
\bc(p)\phi_1(q)\phi_2(k)\phi_2(h)\phi_2(s)
\biggr\}
+\ldots \,,
\eeeqn
where $s=-p-q-k-h$, contains the parameters
\beeqn
\D_{\mu\nu}^{(\bc 2A\phi_1\phi_2)}(p,q,k,h,s)&=&
\de_{\mu\nu}[\de_{20}+\D_{\irr}^{(\bc 2A\phi_1\phi_2)}(p,q,k,h,s)]+
\tilde\D_{\mu\nu}^{(\bc 2A\phi_1\phi_2)}(p,q,k,h,s)
\,,
\\
\D^{(\bc 3\phi_1\phi_2)}(p,q,k,h,s)&=&\de_{21}+
\D_{\irr}^{(\bc 3\phi_1\phi_2)}(p,q,k,h,s)
\,,
\\
\D^{(\bc \phi_1 3\phi_2)}(p,q,k,h,s)&=&\de_{22}+
\D_{\irr}^{(\bc \phi_1 3\phi_2)}(p,q,k,h,s)\,.
\eeeqn
The conditions defining the $22$ relevant parameters are
$$
\D_{\irr}^{(\bc A)}(0)=\partder{p^2}\D_{\irr}^{(\bc A)}(p)|_{p^2=0}=0\,,
\;\;\;\;\;\;\;\;\;
\D_{\irr}^{(\bc \phi_2)}(0)=\partder{p^2}\D_{\irr}^{(\bc \phi_2)}(p)|_{p^2=0}=0
$$
and
$$
\D_{\irr}^{(\bc \cdots)}(p,...)|_{p_i=0}=0
$$
for the other vertices.

Due to these conditions one can isolate in these vertices a dimensional
function of the momenta thus they are irrelevant.
Similarly the vertices ${\tilde \D}_i$ have the
Lorentz indices carried by momenta in a different way with respect to
their relevant parts and are irrelevant.

\vskip 1 true cm
\noindent
{\Large\bf Appendix D}
\vskip 0.3 true cm
\noindent
In this appendix we perform the fine tuning of the couplings of the
effective action.
The condition $\D^{(2)}=0$ gives
\beeqn
&&\de_1=0\;\;\rightarrow\;\;\s_{\bc\chi_2}= M\,,
\\&&\de_2=0\;\;\rightarrow\;\;\s_\alpha= \frac 1 \alpha
+M(\frac{\p\S^{(\bc\chi_2)}}{\p p^2}|_{p^2=0}
+i\frac{\p\S^{(\phi_2 A)}}{\p p^2}|_{p^2=0})\,,
\\&&\de_3=0\;\;\rightarrow\;\;\s_{m_2}=\alpha M^2\,,
\\&&\de_4=0\;\;\rightarrow\;\;\s_2=1\,.
\eeeqn
The condition $\D^{(3)}=0$ gives
\beeqn
&&\de_5=0\;\;\rightarrow\;\;i\s_{2A\phi_1}- M\s_{A\phi_1\phi_2}'
=ieM+m^2 A(0)\,,
\\&&\de_6=0\;\;\rightarrow\;\;M \s_{\bc\chi_2\phi_1}=eM
-i m^2 B(0)\,,
\\&&\de_7=0\;\;\rightarrow\;\;M\s_{\phi_1 2\phi_2}=-m^2
\s_{\bc\chi_1\phi_2}\,,
\\&&\de_8=0\;\;\rightarrow\;\;\s_{\bc\chi_1\phi_2}+\s_{\phi_1 2\phi_2}
\frac{\p\S^{(\bc\chi_2)}}{\p p^2}|_{p^2=0}
=-e-M C(0)-m^2 E(0)\,,
\\&&\de_9=0\;\;\rightarrow\;\;i\s_{A\phi_1\phi_2}'=
M D(0) -2 M C(0)-\s_{\bc\chi_2\phi_1}-\s_{\phi_1 2\phi_2}
\frac{\p\S^{(\bc\chi_2)}}{\p p^2}|_{p^2=0}-m^2 F(0)
\eeeqn
and $\de_{10}$ is zero due to a consistency condition.

The condition $\D^{(4)}=0$ gives
\beeqn
&&\de_{12}=0\;\;\rightarrow\;\;i\s_{4A}=-M I(0)+[A(0)-B(0)]
\s_{2A\phi_1}\,,
\\&&\de_{13}=0\;\;\rightarrow\;\;M\s_{2A2\phi_2}+
\s_{2A\phi_1}\s_{\bc\chi_1\phi_2}=0\,,
\\&&\de_{14}=0\;\;\rightarrow\;\;i\s_{2A2\phi_1}=
\s_{3\phi_1} A(0)+(\s_{A\phi_1\phi_2}'+ie)\s_{\bc\chi_2\phi_1}
+M J(0)\,,
\\&&\de_{15}=0\;\;\rightarrow\;\;
\s_{3\phi_1} B(0)=(\s_{A\phi_1\phi_2}'-ie)\s_{\bc\chi_2\phi_1}
-iM \G^{(\bc\chi_22\phi_1)}(p_i=0)
-M K(0)\,,
\\&&\de_{18}=0\;\;\rightarrow\;\;
M\s_{4\phi_2} +3\s_{\phi_12\phi_2}\s_{\bc\chi_2\phi_1}=0\,,
\\&&\de_{19}=0\;\;\rightarrow\;\;
M\s_{2\phi_12\phi_2}+\s_{3\phi_1}\s_{\bc\chi_1\phi_2}+
2\s_{\phi_12\phi_2}\s_{\bc\chi_2\phi_1}=0
\eeeqn
and $\de_{11}$, $\de_{16}$ and $\de_{17}$ are zero due to
consistency conditions.

The condition $\D^{(5)}=0$ gives
\beeqn
&&\de_{21}=0\;\;\rightarrow\;\;
\s_{4\phi_1}\s_{\bc\chi_1\phi_2}=
-3\s_{2\phi_12\phi_2}\s_{\bc\chi_2\phi_1}
-3 m^2 \G^{(\bc\chi_12\phi_1\phi_2)}(p_i=0)
\\&&\;\;\;\;\;\;\;\;\;\;\;\;
-3 \s_{3\phi_1} \G^{(\bc\chi_1\phi_1\phi_2)}(p_i=0)
-3 \s_{\phi_12\phi_2} \G^{(\bc\chi_22\phi_1)}(p_i=0)
-M \G^{(3\phi_12\phi_2)}(p_i=0)\,,
\eeeqn
and $\de_{20}$ and $\de_{22}$ are zero due to
consistency conditions.

The above solutions are written in terms of some form factors of the
following irrelevant vertices
$$
\G_\mu^{(\bc\chi_1A)}(p,q,k)=p_\mu A+k_\mu B\,,\;\;\;\;\;\;
\S^{(\phi_12\phi_2)}(p,q,k)=(q^2+k^2) C+(qk) D\,,
$$
$$
\S^{(\bc\chi_1\phi_2)}(p,q,k)=q^2 E+k^2 F+(qk) G\,,\;\;\;\;\;\;
\G_\mu^{(2\phi_1\phi_2A)}(p,q,k,h)=p_\mu J+q_\mu K\,,
$$
$$
\G_{\mu\nu\rho}^{(\phi_23A)}(p,q,k,h)=
(q_\mu \de_{\nu\rho}+k_\nu \de_{\mu\rho}+h_\rho \de_{\mu\nu}) H
+[(k+h)_\mu \de_{\nu\rho}+(q+k)_\rho \de_{\mu\nu}+
(q+h)_\nu \de_{\mu\rho}] I
$$
$$+\tilde\G_{\mu\nu\rho}^{(\phi_23A)}(p,q,k,h)\,.
$$
At tree level one has
$$
\s_{m_A}=M^2\,,\;\;\;\;\;
\s_\alpha=\frac {1} {\alpha}\,,\;\;\;\;\;
\s_A=1-\frac {1} {\alpha}\,,
$$
$$
\s_{m_1}=m^2\,,\;\;\;\;\;\s_1=1\,,\;\;\;\;\;
\s_{m_2}=\alpha M^2\,,\;\;\;\;\;\s_2=1\,,
$$
$$
\s_{\phi_2A}=0\,,\;\;\;\;\;\s_{\bc\chi_2}=M\,,
$$
$$
\s_{2A\phi_1}=2Me\,,\;\;\;\;\;
\s_{A\phi_1\phi_2}=-ie\,,\;\;\;\;\;
\s'_{A\phi_1\phi_2}=ie\,,
$$
$$
\s_{3\phi_1}=3!M\frac g e\,,\;\;\;\;\;
\s_{\phi_1 2\phi_2}=2M\frac g e\,,
$$
$$
\s_{\bc\chi_1\phi_2}=-e\,,\;\;\;\;\;
\s_{\bc\chi_2\phi_1}=e\,,
$$
$$
\s_{4\phi_1}=3! g \,,\;\;\;\;\;
\s_{4\phi_2}=3! g\,,\;\;\;\;\;
\s_{2\phi_1 2\phi_2}=2 g\,,\;\;\;\;\;
$$
$$
\s_{2\phi_1 2A}=2e^2\,,\;\;\;\;\;
\s_{2\phi_2 2A}=2e^2\,,\;\;\;\;\;
\s_{4A}=0\,.
$$

\end{document}